\documentclass[aps,preprint,groupedaddress]{revtex4-1}

\usepackage{amssymb}
\usepackage{graphicx}
\usepackage{subfig}
\usepackage{color}
\usepackage{hyperref}

\begin{document}


\title{An agent-based model for emotion contagion and competition in online social media}


\author{Rui Fan$^1$, Ke Xu$^1$ and Jichang Zhao$^{2,\star}$ }
\affiliation{$^1$State Key Lab of Software Development Environment, Beihang University\\
$^2$School of Economics and Management, Beihang University\\
$^\star$Corresponding author: jichang@buaa.edu.cn}


\date{\today}

\begin{abstract}
Recent studies suggest that human emotions diffuse in not only real-world communities but also online social media. More and more mechanisms beyond emotion contagion are revealed, including emotion correlations which indicate their influence and the coupling of emotion diffusion and network structure such as tie strength. Besides, different emotions might even compete in shaping the public opinion. However, a comprehensive model that considers up-to-date findings to replicate the patterns of emotion contagion in online social media is still missing. In this paper, to bridge this vital gap, we propose an agent-based emotion contagion model which combines features of emotion influence and tie strength preference in the dissemination process. The simulation results indicate that anger-dominated users have higher vitality than joy-dominated ones, and anger prefers weaker ties than joy in diffusion, which could make it easier to spread between online groups. Moreover, anger's high influence makes it competitive and easily to dominate the community, especially when negative public events occur. It is also surprisingly revealed that as the ratio of anger approaches joy with a gap less than 10\%, angry tweets and users will eventually dominate the online social media and arrives the collective outrage in the cyber space. The critical gap disclosed here can be indeed warning signals at early stages for outrage controlling in online social media. All the parameters of the presented model can be easily estimated from the empirical observations and their values from historical data could help reproduce the emotion contagion of different circumstances. Our model would shed lights on the study of multiple issues like forecasting of emotion contagion in terms of computer simulations.
\end{abstract}


\maketitle

\section{Introduction}
\label{sub:intro}

Recent years have witnessed the tremendous development of online social media. For example, Twitter attracts more than 310 million monthly active users and produces 500 million tweets every day. In online social media, the high-dimensional user-contributed content, which is a ``big data'' window~\cite{Bliss2012388}, are easily to be collected and analyzed. Because of this, the growth of online social media provides an unprecedented opportunity to study human behavior with unparalleled richness and granularity.

Emotion contagion on real-world social networks is a traditional research area that has been studied by psychologist and sociologist for many years~\cite{hatfield1993emotional_contagion, sigal2002ripple, Fowlera2338dynamic, rosenquist2011social}. After the development of Internet, especially the explosion of online social media, researchers found that users' emotions can also transfer to others via virtual online connections~\cite{bollen2011happiness, guillory2011upset, coviello2014detecting, kramer2012spread, dang2012impact, chmiel2011collective}. Kramer et al. found that sentimental states can be transferred to others via emotion contagion through a massive experiment on Facebook~\cite{Kramer17062014facebook}. Ferrara et al. analyzed the emotion valence of messages that Twitter users receive before they post positive or negative tweets to highlight the effect of emotion contagion~\cite{ferrara2015measuring}. However, a model that simulate the emotion contagion and investigate the diffusion patterns in online social media is still missing. Aiming at filling this gap, in this paper, we propose an agent-based model to study the mechanisms of online emotion diffusion. In previous studies, users' emotions are simply classified to positive and negative or just a score of happiness, neglecting the detailed aspect of human sentiments. However, fine-gained emotion states such as anger or sadness play essential roles in the bursts of social events like earthquake or terrorist attack. In this paper, we categorize human emotions into four categories which are anger, disgust, joy and sadness~\cite{zhao2012moodlens, fan2014anger} to systematically investigate and model the detailed mechanisms of different emotion contagion patterns.

On social networks, emotions flow with the diffusion of information, while information diffusion procedure is also influenced by user emotions behind the contents. Previous studies found that emotional tweets are easier to be retweeted and thus spread more quickly than neutral ones~\cite{stefan2013emotions} and both positive and negative emotions might diffuse in the subsequent comments to corresponding blog entries~\cite{linh2012impact}. In psychology view, each emotion has three dimensions which are valence (positive--negative), arousal (passive/calm--active/excited), and tension (tense--relaxed)~\cite{wundt1907outlines}, among which both valence and arousal impact the diffusion procedure. High arousal emotions such as anger or anxiety boost diffusion more than low arousal emotions such as sadness or contentment~\cite{jonah2011arousal, jonah2012what}. For valence, researchers also found that bad is stronger than good, which implies that negative emotions have more impact than positive ones~\cite{baumeister2001bad}. Kim and Salehan also suggested that negative emotions have higher retweet performance than positive ones~\cite{kim2015effect}. Fan et al. found that anger, which is negative and of high arousal, is more influential than joy while sadness has extremely low correlation between connected users~\cite{fan2014anger}, which are consistent with the findings on the influence of valence and arousal in diffusion procedure.

Meanwhile, the dynamics beyond information diffusion can be influenced by the structure of underlying network such as the characteristics of nodes and ties. Since emotion spread is tightly coupled with the information flow in terms of retweets or comments, its contagion patterns are accordingly affected by the network features. Robert et al. found that nearly all transmission occurred between close friends, suggesting that strong ties are instrumental for spreading both online and real-world behavior in human social networks~\cite{bond201261}. Beyond that, the theory of weak ties, which has been long established, suggests that weak ties enhance diffusion by exchanging novel information across groups~\cite{granovetter1973strength}. Onnela et al. analyzed mobile communication networks and found that removal of the weak ties results in network collapse~\cite{onnela2007structure}. In online social media, weak ties still play a crucial role in the dissemination of information~\cite{Bakshy2012role}. However, Zhao et al. suggested that weak ties can be categorized into positive ones and negative ones which can enhance and prevent the diffusion respectively~\cite{zhao2010weak, zhao2012information}. For the coupling of emotion and tie strength, different emotions would have different preference on propagation ties, e.g., Fan et al. suggested that anger emotion prefers weaker ties compared to joy~\cite{fan2016higher}. In other words, joyful messages tend to be transferred between close friends while angry ones are more likely to evoke resonances of strangers. The existing findings indeed offer insightful mechanisms to establish competent contagion models for emotion in social networks. 

In fact, traditional controlled experiments on emotion diffusion are extremely difficult because of the high costs and spatio-temporal limitations. While with a reliable diffusion model, researchers could study various emotion-related issues such as emotion competition and emotion dynamics in hot-event through computer simulation. However, few existing diffusion models take the influence of emotion and tie-strength preferences into accounts concurrently. In this paper, based on previous works, we proposed an agent-based model which combines the two critical features that function profoundly in the contagion process. Based on the model, we study the tie-strength preference of different emotions, emotion diffusion and how emotions compete in social network. Our model perfectly reproduces the vitality difference and the tie-strength preference of different emotions and offers an comprehensive understanding of emotion contagion and competition, and the emergence of dominated emotion that widely disseminates in the network. Our model would shed lights on study of emotion issues in terms of computer simulations.

\section{Background}
In this section, we introduce our dataset and several up-to-date technologies and findings, which are the foundations of the present work. Our data set is public available and it can be downloaded freely through \url{https://doi.org/10.6084/m9.figshare.4311920.v2}. 

\subsection{Dataset}
Although we concentrate on the emotion contagion model, several experimental results are still necessary to define and estimate the parameters of the model. We collect over 11 million tweets posted by approximately 100,000 users over half a year in Weibo (Chinese Twitter) to perform the experiment. The following edges connected these users and construct a directed network denoted as $G(V, E, W)$, where $V$ and $E$ represent all the users and following edges respectively. $W$ is the set of edges' strengths derived from common friends definition (which can be seen in~\ref{tie-strength-preference}). To investigate the contagion mechanisms of fine-gained emotions, we employ a Bayesian classifier introduced in~\cite{zhao2012moodlens} to category tweets to four emotions which are anger, disgust, joy and sadness. The proportion of the four emotions in these tweets are illustrated in Table~\ref{tab:emotion}, which are important parameters in our model.

\subsection{Emotion correlation}
From the view of conventional social theory, homophily is widely disclosed in social networks~\cite{homophily}. After the explosion of online social media, researchers found that users' physiological states are also assortative in online social networks~\cite{Bliss2012388, bollen2011happiness}. Moreover, anger and joy are revealed to be significantly correlated and anger's correlation is even higher than that of joy~\cite{fan2016higher}. However, disgust and sadness possess extremely low correlations. The emotion with higher correlation indicates that it is easier to affect others, which makes it an important feature of the diffusion process. In this paper, we apply the correlation calculation method on the connected users in our data and achieve similar results as in~\cite{fan2016higher}, as can be seen in Table~\ref{tab:emotion}. In our model, we apply the correlation results of four emotions to construct the model. While in the simulations, we only consider anger and joy and treat the diffusion of disgust and sadness as random reposts due to their extremely low influence. 

\begin{table}[htbp]
\centering
\begin{tabular}{lllll}
\hline
\textbf{} & \textbf{Anger} & \textbf{Joy} & \textbf{Disgust} & \textbf{Sadness} \\
\hline
Proportion & 19.2\% & 39.1\% & 13.7\% & 28.0\%\\
Correlation & 0.41 & 0.35 & 0.04 & 0.03\\
\hline
\end{tabular}
\caption{Statistical results of emotion proportions and emotion correlations of anger, joy, disgust and sadness. The number of joyful messages are larger than that of others. Anger and Joy have large correlations which indicate that they are easily to influence other users.}
\label{tab:emotion}
\end{table}

\subsection{Tie-strength preference}
\label{tie-strength-preference}
Besides emotion, tie strength also has notably influence on diffusion. For example, users tend to communicate via strong ties~\cite{onnela2007structure}, but weak ties are also irreplaceable in diffusion due to its ``bridge'' function between different groups. In this paper, we use the following three tie-strength definitions as in~\cite{fan2016higher}:

\begin{itemize}

\item Common friends. More common friends of two users indicate that their relationship is stronger. Because of this, the first metric we apply is the proportion of \emph{common friends}~\cite{onnela2007structure, zhao2010weak}, which is defined as $c_{ij}/(k_i-1+k_j-1-c_{ij})$ for the tie between user $i$ and $j$, where $c_{ij}$ denotes the number of common friends of $i$ and $j$. And $k_i$ and $k_j$ represent the the degree of $i$ and $j$ respectively.

\item Reciprocity. A higher ratio of reciprocity indicates more trust social connections and more significant homophily~\cite{zhu2014influence}. Therefore, we define the proportion of reciprocal links, through which messages of a specific emotion disseminate, as a measurement of tie-strength of this emotion. Note that the result of reciprocity strength is a proportion rather than an averaged strength as that of the other two definitions.

\item Retweets. The idea of retweet strength is that larger number of retweets represents more frequent interactions. Note that, different from the previous two metrics, retweet strength evolves over time. Therefore, we count only the retweets that occurred before the relevant emotional retweet. Moreover, to smooth the comparison between anger and joy, the retweet strength (denoted as $S$) is normalized by $(S-S_{min})/(S_{max}-S_{min})$, in which $S_{min}$ and $S_{max}$ separately represent the minimum and maximum values of all observations.

\end{itemize}

Based on these three metrics, researchers found that different emotions have different edge preferences in dissemination procedure~\cite{fan2016higher}, i.e., anger prefers weaker ties than joy. People tend to share joyful messages with their close friends, but can be stimulated by strangers' anger in online social media. In this paper, we study the tie-strength preference by the simulation of our model and achieve similar results. Moreover, we reveal the underlying reason of different tie-strength preference of the two emotions.

\subsection{Information diffusion and emotion competition}
Information diffusion is a traditional research area and several classical information diffusion models, such as linear threshold model (LT) and independent cascade model (IC)~\cite{mark1978threshold, Cchen2009efficient, kempe2003maximizing} are well-developed in previous literatures. LT model considers the weight of connection $w_{yv}$ between $y$ and $v$ and generates threshold $\theta_v$ for node $v$. $v$ becomes active if $\Sigma_yw_{yv} \geq \theta_v$, which means the sum of weights between $v$ and its active neighbors exceeds a threshold. With the development of online social media, researchers found that information diffusion presented quite different mechanisms on this platform because of its unique pushing and republishing features~\cite{ICWSM2009gesunheit, zhao2010weak}. Meanwhile, network structure such as tie strength makes significant effect on the diffusion procedure~\cite{onnela2007structure}. Zhao. et al. proposed an information diffusion model which considers the pushing and republishing features and the tie-strength to investigate the relationship between the tie strength and information propagation in online social networks~\cite{zhao2012information}. Our model absorbs the advantages of previous literatures such as the pushing and republishing mechanisms of online social media, the preference of tie strength in diffusion procedure, and thresholds which are defined to determine the condition of diffusion.

As the theory of economics of attention~\cite{simon1971designing,crawford2015world,davenport2013attention, josef2007attention}, users' attention is a scarce resource that multiple information competes for. With the development of Internet, especially the explosion of online social media, netizens receive redundant messages from various online channels such as Facebook, Twitter and online forums, which results in extremely severe information overload problem. Because of this, information competes for users' finite attention and the winners go vital in cyber space~\cite{michael1997attention, Wu06112007}. To investigate the competition mechanisms, researchers also proposed competition models~\cite{moussaid2009individual,weng2012competition,sneppen2010minimal,karrer2011competing}, e.g., Weng et al. proposed an agent-based model to study memes competition in Twitter. However, except for hashtags, topics and memes, emotions also compete for users' limited attention and the competitive emotion would pervasively diffuse in the network. Based on this idea, we introduce a structure in our model to represent users' limited attention and analyze the diffusion results from the perspective of emotion competition.

\section{Agent-based model}
\begin{figure}[htb]
\centering
\includegraphics[width=0.8\linewidth]{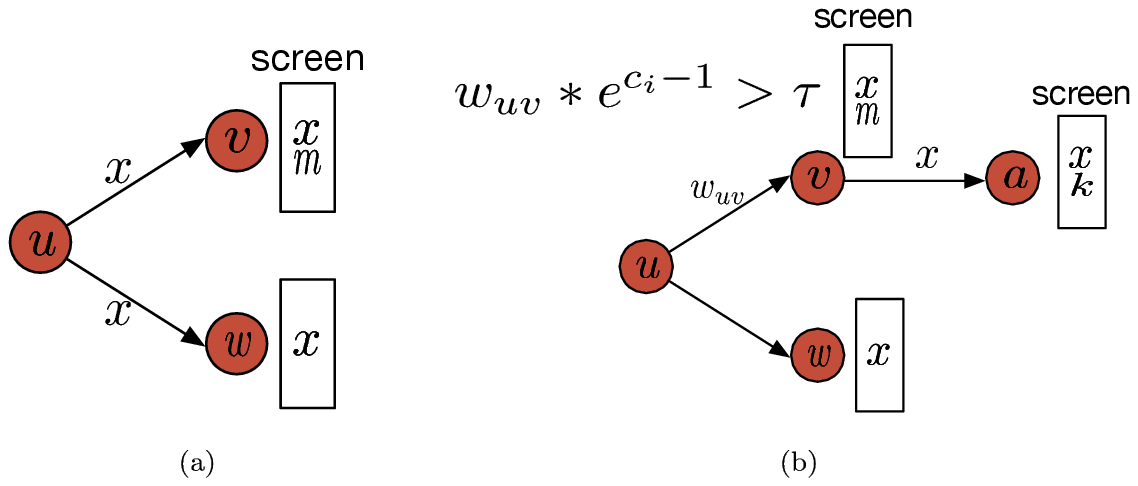}
\caption{Pushing and republishing procedures in our model. In (a), $v$ and $w$ follow $u$. When $u$ posts a new tweet $x$ with probability $P_n$, $x$ is put into the top of $v$ and $w$'s screens. (b) illustrates the republishing process. $v$ reads $x$ that in his/her screen, if the retweet tendency $w_{uv}*e^{c_i-1}>\tau$ ($c_i$ is the correlation of the emotion $i$ delivered by $x$) is higher than $\tau$, $x$ will be reposted by $v$ and enter the screen of $a$, who is the follower of $v$, implying the emotion $i$ initially posted by $u$ transfers from $v$ to $a$.}
\label{fig:pushing_republishing}
\end{figure}

Agent-based model is widely employed in previous works to study network community, discussion process, emotion expression patterns and meme competition~\cite{Schweitzer2010, Ding2010, Sobkowicz2010, boost_forum, weng2012competition}. Inspired by the model introduced in~\cite{weng2012competition}, we develop an emotion contagion model that combines emotion influence and tie strengths, to study emotion contagion mechanisms in social network. We simulate the diffusion process on the real-world following network $G(V, E, W)$ and diffusion behaviors are reflected in terms of retweets which occur on connected edges. Each user has a screen which contains $N$ time-ordered messages that he/she has received. If users he/she follows post or repost messages, these messages will be pushed into the screen. In one circulation, a randomly selected user posts new messages with probability $P_n$, or reposts several ones that in the screen. If the number of messages in the screen exceeds $N$, the old messages will be removed. This process stems from the pushing and republishing mechanisms in online social media that users either post new messages or read messages he/she received and repost some of them. The critical principle of the dissemination is that if one message with emotion of significant influence and the tie strength between the two users are strong, the message will be preferentially reposted. The model is defined as follows:

\begin{itemize}
\item Step 1: Randomly select a user $u$ from $V$.

\item Step 2: $u$ posts a new message $m$ with the probability $P_n$. The emotion of $m$ is decided by the statistical results of empirical data on which the probabilities (denoted as $P_{anger}, P_{disgust}, P_{joy}$ and $P_{sadness}$) of anger, disgust, joy and sadness are 19.2\%, 13.7\%, 39.1\% and 28.0\%, respectively, as can be seen in Table~\ref{tab:emotion}. At last, $m$ is pushed to the screens of all followers of $u$. If $u$'s screen contains more than $N$ tweets, oldest tweets will be removed to narrow the screen until its size is $N$. The procedure of pushing new messages is illustrated in Fig.~\ref{fig:pushing_republishing}(a).

\item Step 3: If $u$ does not post new message, he/she will read all the messages in his/her screen. When he/she reads a message that has emotion $i$, and the tie strength of $u$ and his/her followee who posts or reposts this message is $s$, the retweet tendency is defined as $t=s*e^{c_i-1}$, where $c_i$ is the correlation of emotion $i$. If the tendency $t$ exceeds a threshold $\tau$, the message will be reposted and pushed to the screens of all followers of $u$. We also remove oldest tweets that in $u$'s screen if the screen overloads. We call the above procedure republishing, which is shown in Fig.~\ref{fig:pushing_republishing}(b).

\item Repeat the above three steps for $M$ times.
\end{itemize}

Table~\ref{tab:params} illustrates important parameters in our model. $P_{emotion}$ such as $P_{anger}$ and $P_{joy}$ is the probability of four newly posted emotions, which can be adjusted to reflect the variation of new messages with one dominant emotion in social events. $w_{uv}$ is achieved from the common friends strength definition, representing the strength of users $u$ and $v$. $N$ represent users' screen size, which is set to 20 in this paper because Weibo presents 10 newest tweets per page in their App. It can further be adjusted to investigate emotion contagion in different attention spans. After removing isolated nodes, we achieve 88,532 connected users which are used to construct $G$. Because of this, the simulation steps $M$ is arbitrarily set to 8,853,200, which is 10 times of the network size and supposed to be sufficient in simulating the diffusion. We tune the threshold $\tau$ in the following simulation and further estimate the threshold in real online social media.

\begin{table}[htbp]
\centering
\begin{tabular}{lll}
\hline
\textbf{Symbol} & \textbf{Meaning} & \textbf{Value} \\
\hline
$P_n$ & The probability of posting new tweet & 0.45\\
$P_{emotion}$ & The probability of posting tweet with one emotion, such as $P_{anger}$ & See Table~\ref{tab:emotion}\\
$c_{emotion}$ & The correlations of the four emotions & See Table~\ref{tab:emotion} \\
$w_{uv}$ & The tie-strength between $u$ and $v$ & -- \\
$N$ & The size of the screen & 20 \\
$M$ & The total number of steps & 8,853,200 \\
$\tau$ & The retweet threshold & -- \\
\hline
\end{tabular}
\caption{The description of parameters of the model.}
\label{tab:params}
\end{table}

Our model considers two important features: emotion correlation and the tie-strength. Emotion correlation represents emotions' influence. Hence emotion with high correlation such as anger is easier to be reposted. Meanwhile, as described in~\cite{onnela2007structure}, users with stronger connections have more communications. Our model simulates the diffusion based on these two findings as described in Step 3: if the product of tie strength $s$ and $e^{c_i-1}$ is larger than a threshold, the message will be reposted. This definition combines the influence of content that is presented by emotion and the the network structure which is presented by tie-strength. The code of the model is also publicly available and it can be freely downloaded through \url{https://doi.org/10.6084/m9.figshare.5092522.v1}. 

\section{Results}
In this section, we first report the results of our emotion contagion model, including tie-strength preferences and user vitality difference of different emotions, estimation of parameters $\tau$ by comparing the simulation with empirical results, emotion diffusion and competition mechanisms.

\subsection{Tie-strength preference}

\begin{figure}[htb]
\centering
\includegraphics[width=0.8\linewidth]{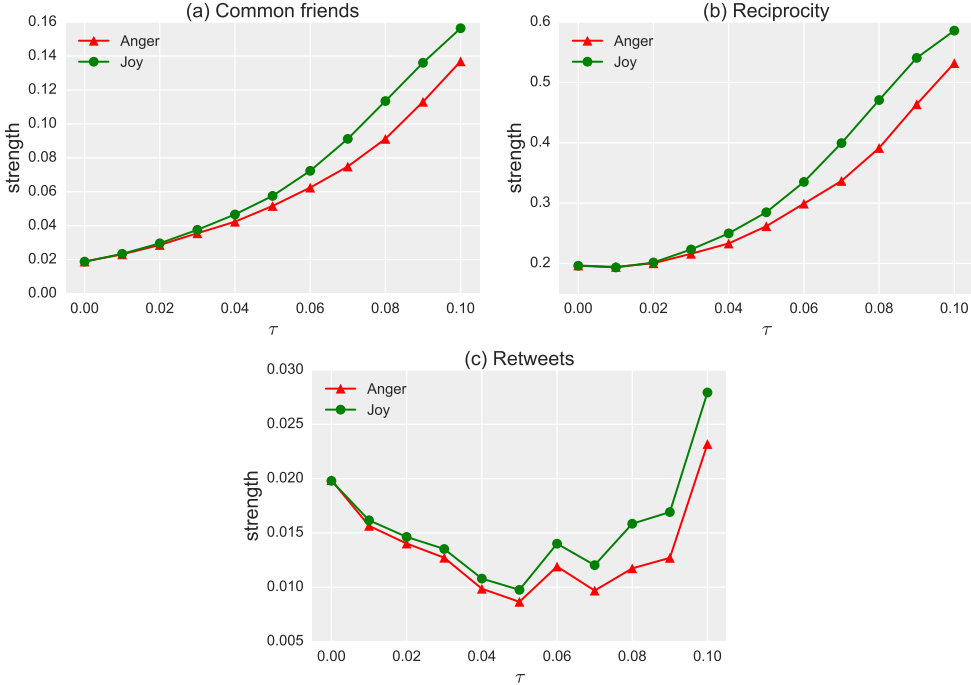}
\caption{Tie-strength preferences of anger and joy. In all the three strength definitions, including common friends (a), reciprocity (b) and retweets (c), anger prefers weaker ties than joy, especially when $\tau$ is large.}
\label{fig:strength}
\end{figure}

Based on the three strength definitions, we calculate the average strength of ties on which anger and joy emotion transfer. Note that given the extremely low influence of disgust and sadness, which can be learnt from their poor corelations in the social network as can be seen in Table~\ref{tab:emotion}, only anger and joy are explored in the following simulations of emotion contagion. As can be seen in Fig.~\ref{fig:strength}, all the results of the three strength definitions indicate that anger prefers transferring through weaker ties than joy. When the diffusion threshold $\tau$ is very small, e.g., $\tau=0$, all messages can be retweeted regardless of the emotion and tie-strength. As a result, both of anger and joy diffuse on edges that have similar average strength. With the growth of $\tau$, edges with small strength prevent some messages spreading on them, especially the ones with small correlations. On some edges, joyful messages cannot spread while angry ones could still pass. As a result, the average tie-strength for contagion of both anger and joy grows up with the growth of $\tau$ while their difference increases accordingly. Classical social theory suggests that weak ties play a key role in diffusion process because they always connect different communities and transfer information from one group to another. Because of this, the preference of weaker ties could make anger easier to spread widely. The preferences of tie strengths unraveled by our model are consistent with the empirical results in~\cite{fan2016higher}, which also suggested that anger prefers weaker ties than joy. Moreover, our model reveals where the different preferences come from. The diffusion in the network depends on the tie-strength and the influence of the contents' emotion simultaneously. More influential emotions such as anger do not rely on ties with large strengths too much, which makes it easily spread via weaker ties than joy.

\begin{figure}[htb]
\centering
\includegraphics[width=0.8\linewidth]{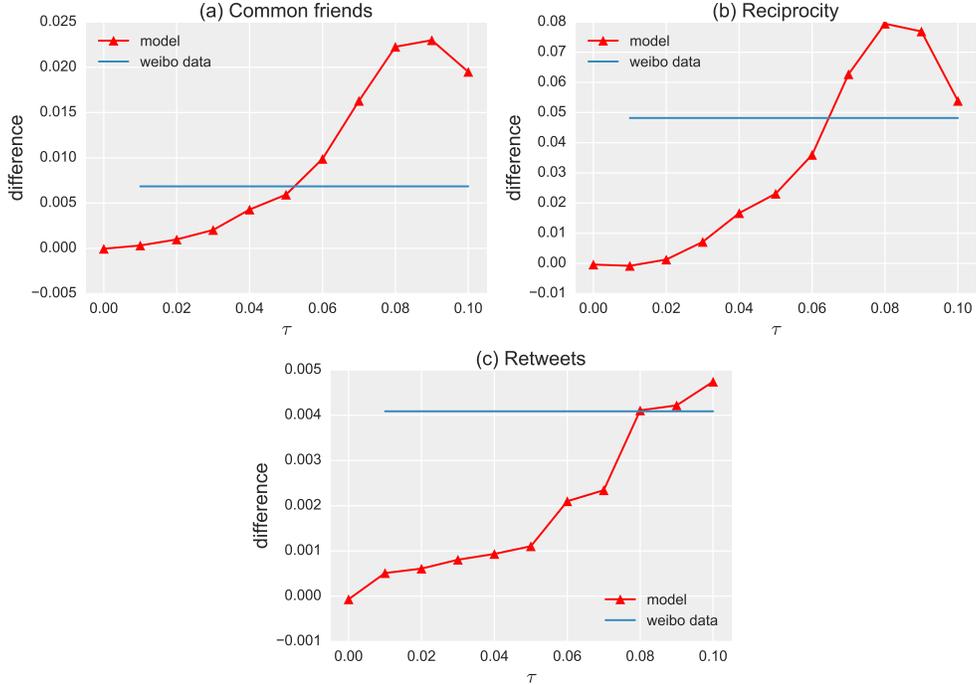}
\caption{The tie-strength difference of joy and anger. The difference is defined as $\bar s_{joy}-\bar s_{anger}$, where $\bar s_{anger}$ and $\bar s_{joy}$ are the average diffusion strengths of anger and joy respectively. The horizontal lines represent the difference calculated from realistic Weibo data~\cite{fan2016higher}.}
\label{fig:diff}
\end{figure}

Moreover, we estimate the value of $\tau$ based on empirical analysis and the results are illustrated in Fig.~\ref{fig:diff}. The empirical results of strength preference comes from~\cite{fan2016higher}, which are calculated from six months tweets posted by the network users based on the three strength definitions. As can be seen, we can draw the conclusion that the value of $\tau$ in real online social media ranges from 0.05 to 0.08. Our model results are assortative with observations from the real-world data and we can further estimate the threshold in Weibo, which are strong evidence of the reliability of our model.

\subsection{User Vitality}
\begin{figure}[htb]
\centering
\includegraphics[width=0.8\linewidth]{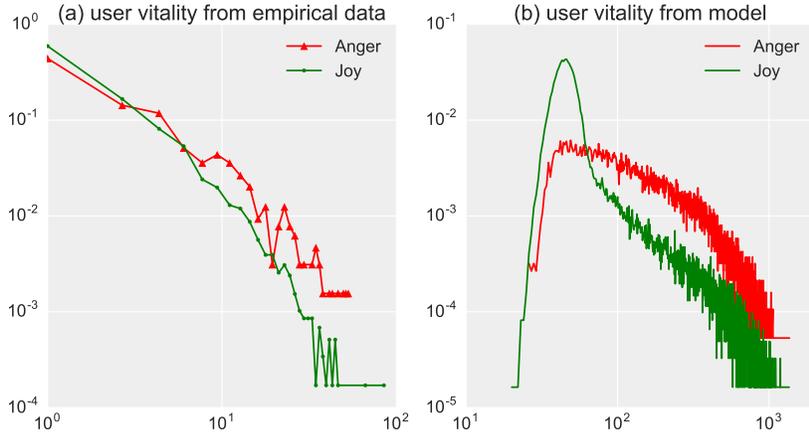}
\caption{The distributions of user vitality for respectively anger and joy-dominated users. (a) User vitality from the empirical data. (b) User vitality in simulations and here we set $\tau=0.06$, which is close to empirical circumstance, as can be seen in Fig.~\ref{fig:diff}. It is worth noting that different from the empirical data, there is an increment in vitality distribution of the model as the user vitality is very low. It comes from the configuration of the model that agents are evenly selected to publish or republish tweets at each step, which results in relatively less low-vitality users as compared to the empirical data.}
\label{fig:activity}
\end{figure}

The user vitality, which reflects the extent of a user’s activeness in online social media, can be an excellent indicator of users' intention in participating online activities. Here the user vitality is intuitively defined as the number of messages posted per day by a user after the registration in the empirical data. And for the simulations in the model, since all agents share the same age, the user-activity can be directly valued by the total number of posted messages by an agent. Then besides strength preferences, it is surprisingly found that in both simulation and empirical data, anger-dominated users demonstrate higher vitality than their joy-dominated counterparts, as can be seen in Fig.~\ref{fig:activity}, in which anger (joy)-dominated users are those posting more angry (joyful) tweets than that of other emotions. The similarly long-tailed distribution of user vitality reproduced by our model, as shown in Fig.~\ref{fig:activity}, again indicates its effectiveness in replicating emotion contagion in social media. And in fact, the consistence between the empirical data and our model further implies that being a sentiment with high arousal, anger can indeed enhance users' activeness in online activities and make itself competent in capturing attention during the contagion. 

\subsection{Emotion diffusion and competition}
\begin{figure}[htb]
\centering
\includegraphics[width=0.8\linewidth]{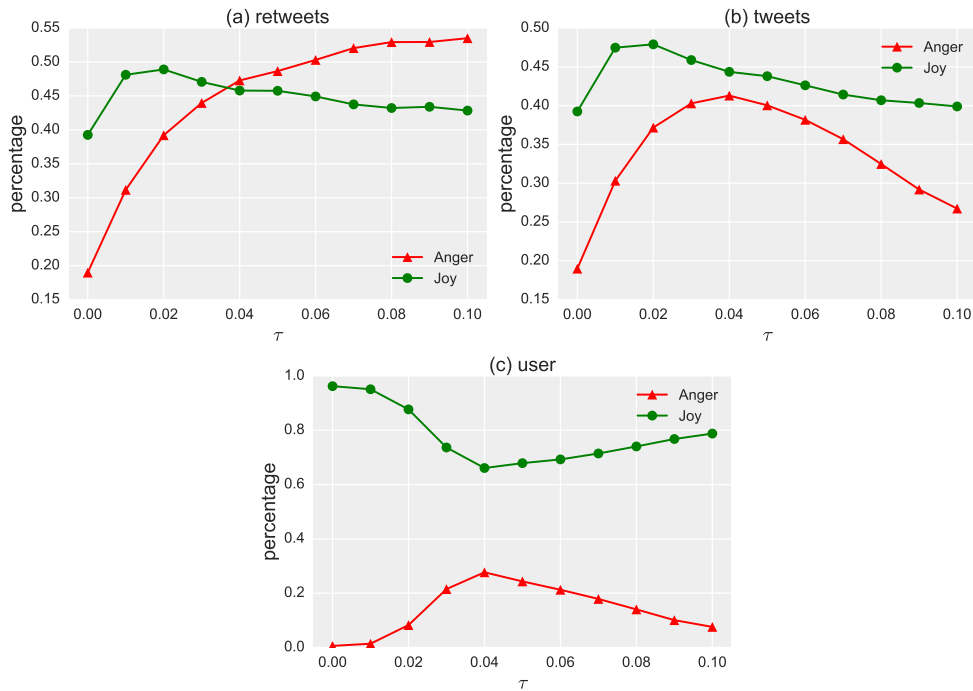}
\caption{The simulation results of anger and joy, in which emotion probabilities of new messages are 19.2\%, 13.7\%, 39.1\% and 28.0\% for anger, disgust, joy and sadness, respectively. These proportions are calculated from real Weibo data. In all of the three figures, x-axises represent the threshold $\tau$ and y-axises are the percents of retweets, tweets and users after the simulation. (a) and (b) separately illustrate the percentages of anger and joy in reposted messages and all messages including new tweets and reposted tweets. (c) is the proportion of anger- and joy-dominated users. Users who post or repost more angry (joyful) tweets than other emotions will be similarly labeled as anger(joy)-dominated users. And the number of a specific emotion dominated users reflects the diffusion scope of this emotion.}
\label{fig:percents}
\end{figure}

In this section, we discuss the emotion contagion patterns based on the simulation results. Fig.~\ref{fig:percents}(a) illustrates the retweets proportions of anger and joy. When $\tau=0$, tweets with all of the four emotions can be transferred via all edges and accordingly the retweets proportions of anger and joy are assortative with $P_{anger}$ and $P_{joy}$, respectively. While with the growth of $\tau$, angry tweets can spread on some weak ties on which joyful tweets can't transfer and as a result, the proportion of anger retweets exceeds that of joy when $\tau>0.03$. However, the proportions of all tweets decline because the number of retweets will decline when the threshold is large, as can be seen in Fig.~\ref{fig:percents}(b). And the proportions of anger and joy are closest when $\tau=0.04$ but the number of angry tweets is still below joy because of the smaller value of $P_{anger}$. Anger-dominated users also reach a peak when $\tau=0.04$. Because of the higher influence of anger, its number of retweets increases with the rising of $\tau$ and the total angry tweets and anger-dominated users also reach peaks at specific $\tau$. However, as $P_{anger}$ is much lower than $P_{joy}$, the total numbers of angry tweets and the anger dominated users are smaller than that of joy for all $\tau$. From the view of emotion competition, people's finite attention are filled with messages with different emotions. When the number of messages with one emotion increases, it becomes more competitive. Two factors that influence the amount of messages with one emotion are the probability of newly posted emotion messages and the amount of republishing messages with this emotion because both of pushing and republishing tweets will appear in followers' screens. According to Weibo data, $P_{joy}$ is much larger than $P_{anger}$, while anger has higher influence and easy to be republished which produces more angry retweets. In conclusion, anger's higher influence can enhance its competition and makes it spread extensively from both the perspectives of messages and users.

\begin{figure}[htb]
\centering
\includegraphics[width=0.8\linewidth]{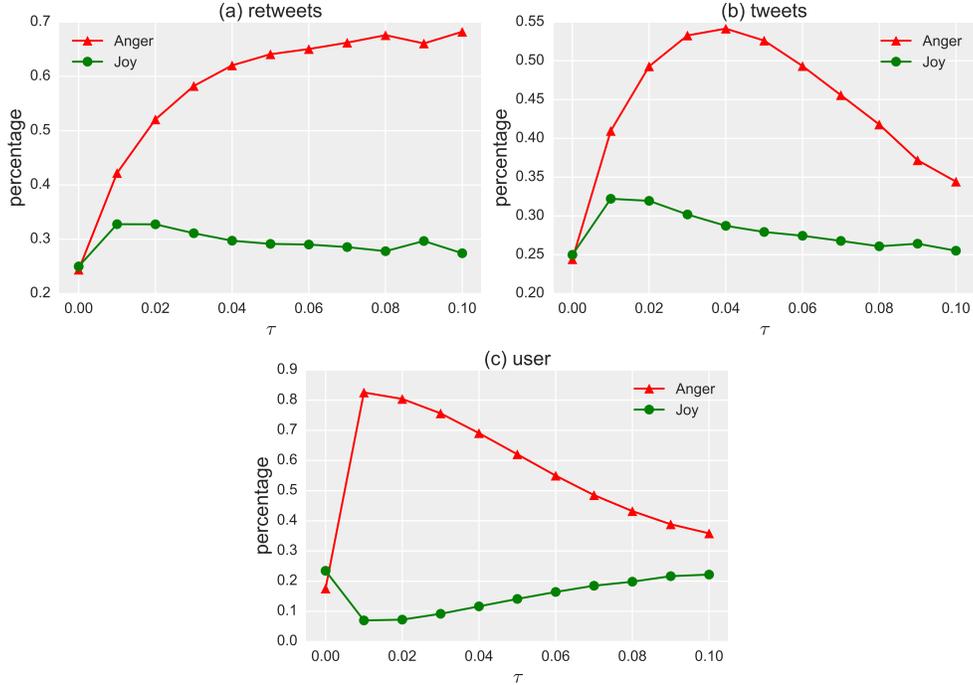}
\caption{Simulation results of emotion competition. New tweets have the same probabilities (25\%) to be labeled to the four emotions. (a) illustrates the percentages of angry and joyful retweets. The proportion of angry retweets increases with the growth of $\tau$ while of joy, the value first increases to a peak and then declines. (b) is the percentages of total tweets including new tweets and reposted tweets. The proportions of both of anger and joy increase first and then begin to drop. And the values of anger far exceed that of joy. (c) represents the anger and joy dominated users and the values of anger are still much more than that of joy.}
\label{fig:cmpt}
\end{figure}

Moreover, in some circumstances, especially when negative societal events occur, people will post many angry messages in online social media and the possibility of publishing angry tweets will grow significantly. Because of this, we tune the emotion probabilities of newly posted tweets to 25\% for all the four emotions to study the emotion competition from a equitable start. As can be seen in Fig.~\ref{fig:cmpt}(b) and Fig.~\ref{fig:cmpt}(c), the proportions of angry tweets and anger-dominated users are much more than that of joy messages. When $\tau=0$, the proportions of angry and joyful tweets and users are similar. With $\tau$ grows up, the proportions of anger surpass joy and their difference in coverage reach a maximum at specific $\tau$, which is about 0.04 for tweets and 0.01 for users respectively. However, when $\tau$ is too large, tweets with both anger and joy are hard to spread in the network, leading to a decreasing gap between them. In this simulation, joy has no superiority in the newly posted messages, and anger still has more influence. As a result, anger could dominate users' attention, which makes it more competitive than joy. In conclusion, when negative events occur, people will post many angry tweets which are easily to influence others. As a result, angry users will dominate the whole network. It can well explain the bursty rise of negative events, in which anger dominates the public opinion~\cite{fan2016higher}. Because of anger's high influence, website managers should try to prevent its extensively and unreasonably spread from the very beginning.

\begin{figure}[htb]
\centering
\includegraphics[width=0.8\linewidth]{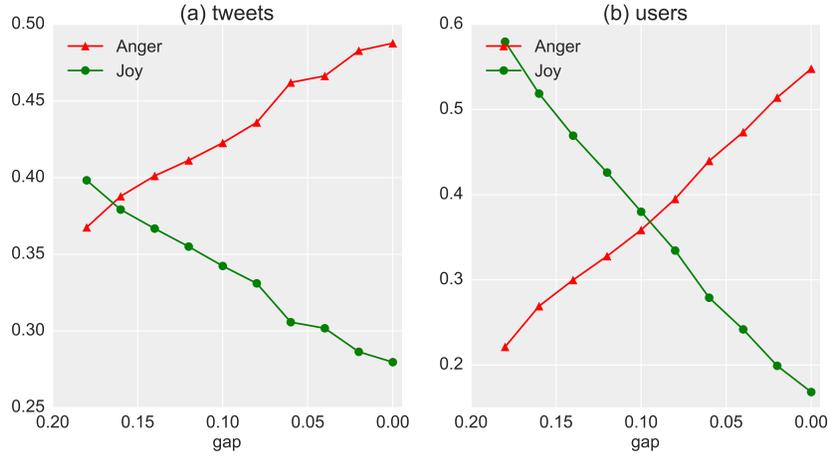}
\caption{The proportions of angry and joy messages and anger and joy-dominated users with the reduction of $P_{joy}-P_{anger}$. In this figure, we fix $P_{disgust}$ and $P_{sadness}$ to 0.25 and adjust $P_{joy}-P_{anger}$. For example, when $P_{joy}=0.30$ and $P_{anger}=0.20$, the gap of them is 0.10. (a) illustrate that when the gap is smaller than 0.16, the proportion of angry messages surpasses that of joy messages. (b) illustrate the proportion of anger and joy-dominated users. When the gap is smaller than 0.10, the proportion of anger-dominated users exceeds that of joy-dominated users.}
\label{fig:exceed}
\end{figure}

To find the critical point that anger surpasses joy in the network, we tune the probability of newly posted anger and joy messages, denoted as $P_{anger}$ and $P_{joy}$, to investigate when anger could dominate the network as $P_{anger}$ approaches $P_{joy}$. As can be seen in Fig.~\ref{fig:exceed}, when the gap is large, e.g., $P_{joy}-P_{anger}=0.2$, both of joyful tweets and joy-dominated users are more than angry ones. With shrinking of the gap, the number of both angry messages and anger-dominated users increases, and when the gap is smaller than 0.16 and 0.10, the proportion of angry messages and anger-dominated users exceeds that of joy messages and joy-dominated users respectively. The different critical gaps for tweets and users comes from the vitality difference, i.e., angry users demonstrate higher activeness in online activities. As a result, the number of angry messages exceeds that of joy although the number of anger-dominated users are still smaller than that of joy-dominated users. This finding demonstrate that when the proportion of newly posted angry messages is 10\% less than that of joy, the network could be dominated by angry users. Network managers should be aware of that and adopt corresponding techniques to prevent the collective outrage in cyber space.

\section{Conclusion}
The explosive development of online social network produces massive data which provides an unprecedented opportunity to investigate human behavior. In this paper, we propose an agent-based model to well simulate the emotion contagion in online social media. Our model absorbs advantages by considering the pushing and republishing of online social media, the threshold which is the diffusion condition and the competition of users' limited attention. Meanwhile, the most innovative point of our model is that we combine the influence of content which is presented by emotion and the network feature which is tie-strength. The primary idea of this combination is that higher emotion influence and stronger ties enhance diffusion in social network. By analyzing the simulation result, we find that as compared to joy, angry users possess higher vitalities and anger can spread on weaker ties which means it could widely diffuse across communities. These findings are consistent with previous empirical observations on Weibo. But our model further disclose the origin of this phenomenon which is high influence makes anger easily transfer via weak ties. Our model can also be used to study the emotion contagion, e.g., from the simulation results, we find that higher influence of anger can enhance its competition. Moreover, when the probability of newly posted angry tweet is high, i.e., in circumstances of negative societal events, anger will influence massive users and even dominate the network due to its higher influence. We also give a guide value that when the proportion of newly posted anger messages is 10\% less than that of the joy, the network could be dominated by negative emotion. This is a warning signal in early stages, i.e., in some societal events, anger emotion could widely propagate and lead to the collective online rage.

Our model demonstrate that anger's high influence can enhance its dissemination, and in some conditions, it can even dominate the network. Network managers should realize anger's capability of being highly contagious and apply techniques to try to prevent the unreasonable prevalence of anger. Our emotion contagion model grabs multiple critical features of online diffusion and can be used to study multiple issues of emotion diffusion and even forecasting by adjusting parameters. For example, we simulate the diffusion of collective social event by changing the probabilities of newly posted emotions. Researchers can also study the influence of intention by adjusting the size of the screen. 

\begin{acknowledgments}
This work was supported by NSFC (Grant No. 71501005) and the fund of the State Key Lab of Software Development Environment (Grant Nos. SKLSDE-2015ZX-05 and SKLSDE-2015ZX-28). R. F. also thanks the Innovation Foundation of BUAA for PhD Graduates.
\end{acknowledgments}


\end{document}